\documentclass[11pt]{article}

\usepackage{
graphicx,amsmath,amssymb,bm}

\newtheorem{prop}{Prop}[section]
\newtheorem{lemma}[prop]{Lemma}
\newtheorem{definition}[prop]{Definition}
\newtheorem{theorem}[prop]{Theorem}

\newtheorem{note}[prop]{Note}
\newtheorem{corollary}[prop]{Corollary}

\title{
Self-averaging of perturbation Hamiltonian density\\
in perturbed spin systems 
}
\author{C. Itoi  \\
Department of Physics, GS and CST, Nihon University, \\
Kanda-Surugadai, Chiyoda, Tokyo 101-8308, Japan} 
\begin{document}
\maketitle
\begin{abstract}{ It is shown that the variance of a perturbation Hamiltonian density
vanishes in the infinite-volume limit  of the perturbed spin systems with quenched disorder.  
This is proven in a simpler way and under less assumptions  than before.
A corollary of this theorem  indicates the impossibility of  non-spontaneous replica symmetry-breaking in disordered spin systems. 
The commutativity  between the infinite-volume limit and the  switched-off  limit of a replica symmetry-breaking perturbation implies that the variance of the spin overlap  vanishes
in the replica symmetric Gibbs state.  
}
\end{abstract}



%
\section{Introduction}
Recently, it was proven that the variance of the perturbation Hamiltonian density vanishes in the infinite-volume limit of  disordered  quantum spin systems 
\cite{I2}.  In the present paper, we give a simpler proof of this theorem under less assumptions on the models. 
First, we give a definition of the model and the main theorem. 
We study quantum spin systems on a finite set 
$V_N$ with $|V_N| =N$. 
Spin operators $S^{p}_j$ $(p=x,y,z)$ at  a site $j \in V_N$ acting
on a Hilbert space ${\cal H} :=\bigotimes_{j \in V_N} {\cal H}_j$ are
defined by a tensor product of the spin matrix acting on ${\cal H}_j \simeq {\mathbb R}^{2S+1}$ and unit operators, where $S$ is an arbitrarily fixed positive semi-definite half integer.
These operators are self-adjoint and satisfy the commutation relations
$$
[S_j^x,S_k^y]=i \delta_{j,k} S_j^z ,  \ \  \ \ \  [S_j^y,S_k^z]=i \delta_{j,k}  S_j^x, \ \ \  \ \ 
[S_j^z,S_k^x]=i \delta_{j,k} S_j^y,
$$
and the spin at each site $i \in V_N$ has a fixed magnitude 
$$
 \sum_{p=x,y,z}(S_j^p)^2 =S(S+1) {\bf 1}.
$$
We define an  unperturbed Hamiltonian $H_N({\bf S})$ first.
 ${\cal P} (V_N)$ denotes the collection of all subsets of $V_N$.
 Let ${\cal C}_N^p \subset {\cal P}(V_N)$ be a collection of interaction ranges which are bounded  subsets of $V_N$ and denote
$$M:=|\cup_{p=x,y,z}  {\cal C}_N^p|.$$ 
 Let ${\bf J} = (J_X^p)_{X \in {\cal C}_N^p, p=x,y,z}$ be a sequence of  real-valued independent random variables 
with finite expectations
$ 
{\mathbb E} J_X^p \in {\mathbb R}$, where ${\mathbb E}$ denotes the expectation over ${\bf J}$. 
Assume the following bound on their variances by a positive constant $\sigma$ independent of  $N$  
 $$
\sum_{p=x,y,z} 
\sum_{X\in{\cal C}_N^p} {\mathbb E} (J_X^p -{\mathbb E J_X^p} )^2 \leq \sigma^2 N.
 $$ 
Denote a sequence of spin operators  
$S_X^p:=(S_j^p)_{j \in X, p=x,y,z}$  on a subset $X$ and let $\varphi_X^p$ be a self-adjoint-operator-valued bounded
 function of $S_X^p$, such that $\| \varphi_X^p(S_X^p) \| \leq C_\varphi$, where
the operator norm is defined by 
$\| O \|^2 :=\sup_{(\phi,\phi)=1 } (O \phi, O \phi)$ for an arbitrary linear operator $O$ on ${\cal H}$.
 We consider a model defined by the following Hamiltonian with ${\cal C}_N^p$, $\varphi_X^p$ and ${\bf J}$
\begin{equation}
H_N({\bf S},{\bf J}) := \sum_{p=x,y,z}\sum_{X\in {\cal C}_N^p} J_X^p \varphi_X^p(  S_X^p). \label{unpert}
\end{equation}
One can assume a symmetry of the Hamiltonian $H_N({\bf S},{\bf J})$, if one is interested in symmetry-breaking phenomena.
To detect a spontaneous symmetry-breaking, the long-range order of 
an order operator  is utilized  in the symmetric Gibbs state.  Although the symmetric Gibbs state with long-range order is mathematically well defined, 
such a state is unstable due to strong fluctuations and it cannot be realized.  
On the other hand,  it is believed  that a perturbed Gibbs state with an infinitesimally 
symmetry-breaking term in the Hamiltonian 
is stable and  realistic.
Let ${\cal O}_N$ be  a symmetry-breaking self-adjoint operator acting on ${\cal H}$ with a uniform bound
\begin{equation}
\| {\cal O}_N \| \leq C_o,  
 \end{equation} 
where $C_o$ is a positive constant independent of the 
system size $N$. 
This operator ${\cal O}_N$ can perturb the model as  a symmetry-breaking perturbation.
At the same time, it's Gibbs expectation can measure the symmetry-breaking as an order parameter.
Define a perturbed Hamiltonian by 
\begin{equation}
H_\lambda:= H_N({\bf S}, {\bf J})-N \lambda {\cal O}_N,
\label{hamil}
\end{equation}
where $\lambda \in {\mathbb R}$.
To study spontaneous symmetry-breaking, one can regard ${\cal O}_N$ as an order operator which breaks the
 symmetry. 
  For instance, ${\cal O}_N$ is a
spin density
 \begin{equation}
{\cal O}_N = \frac{1}{N} \sum_{j \in V_N} S_j^z.
 \end{equation}

Define the Gibbs state with the Hamiltonian (\ref{hamil}).
For  $\beta>0$,  the  partition function is defined by
\begin{equation}
Z_N(\beta, \lambda, {\bf J}) := {\rm Tr} e^{ - \beta  H_N({\bf S}, {\bf J})+\beta N  \lambda {\cal O}_N}
\end{equation}
where the trace is taken over the Hilbert space ${\cal H}$.
Let  $f$ be an arbitrary function 
of spin operators.  
The  expectation of $f$ in the Gibbs state is given by
\begin{equation}
\langle f({\bf S}) \rangle_N,_\lambda=\frac{1}{Z_N(\beta, \lambda, {\bf J})}{\rm Tr} f({\bf S})  e^{ - \beta  H_N({\bf S}, {\bf J})+\beta N  \lambda {\cal O}_N}.
\end{equation}
Define the following function from the partition function
 \begin{equation}
\psi_N(\beta, \lambda,{\bf J}):=\frac{1}{N} \log Z_N(\beta,\lambda, {\bf J}),
 \end{equation}
and its expectation 
 \begin{equation}
p_N(\beta, \lambda):={\mathbb E} \psi_N(\beta, \lambda, {\bf J}).
 \end{equation}
where ${\mathbb E}$ denotes the expectation over ${\bf J}$.
The function $-\frac{N}{\beta} \psi_N$ is called free energy of the sample in statistical physics.\\

In the present paper, we require  the following two assumptions on the Gibbs state defined by the perturbed  Hamiltonian (\ref{hamil}). \\
 
\noindent
{{\bf Assumption 1}
  \it
The  infinite-volume limit  of the function $p_N$ 
 \begin{equation}
 p(\beta, \lambda) = \lim_{N\to\infty}   p_N(\beta, \lambda), 
 \end{equation}
exists for each $(\beta, \lambda) \in (0,\infty) \times {\mathbb R}$.}\\

\noindent
{{\bf Assumption 2} \it  The following commutator of
the perturbation operator ${\cal O}_N$ and the Hamiltonian vanishes in the infinite-volume limit 
 \begin{equation}
\lim_{N\to\infty} \|[{\cal O}_N, [H{(\bf S}, {\bf J}), {\cal O}_N]]\| =0,
 \end{equation}
for an arbitrarily fixed sequence $\bf J$.
}\\

Assumption 1 has been proven several spin models \cite{CL,Gff2,I,KS,L,PF,V,Z}.
Any short-range Hamiltonian and a spin density as a perturbation operator ${\cal O}_N$
 satisfy Assumption 2.
In the present paper, we prove the following main theorem for an arbitrary spin model with 
the Hamiltonian (\ref{hamil})  satisfying Assumptions 1 and 2.  
Though  the proof in Ref. \cite{I2} needs the assumption that the variance of $\psi_N({\bf J})$ vanishes in the infinite-volume limit,
we prove it as a lemma in the present paper.
\\

{\theorem \label{MT} Consider  a quantum spin model defined by the Hamiltonian
(\ref{hamil}) with perturbation operator ${\cal O}_N$ satisfying Assumptions 1 and 2.  The expectation of  the order  operator
 \begin{equation}
 \displaystyle{\lim_{N\to\infty }{\mathbb E}  \langle {\cal O}_N \rangle_N,_{\lambda}}, 
 \end{equation}
 exists  in the infinite-volume limit  for almost all $\lambda$ and 
 its variance in the Gibbs state  and the distribution of disorder vanishes 
 \begin{equation}
\lim_{N \to \infty }{\mathbb E}\langle ( {\cal O}_N -{\mathbb E} \langle {\cal O}_N \rangle_N,_{\lambda} )^2 \rangle_N,_{\lambda}=0,
\end{equation} 
in the infinite-volume limit  for almost all $\lambda \in {\mathbb R}$.}\\

Theorem \ref{MT} implies  also the existence of the following infinite-volume  limit for almost all $\lambda \in {\mathbb R}$
 \begin{equation}
\lim_{N\to\infty}
  {\mathbb E} \langle {\cal O}_N^2 \rangle_N,_{\lambda}=(\lim_{N\to\infty}{\mathbb E} \langle {\cal O}_N\rangle_N,_{\lambda})^2.
 \end{equation}
The perturbation operator ${\cal O}_N$ is self-averaging  in the perturbed model.  
Although Theorem \ref{MT} is physically  natural and physicists believe it by
their experiences supported by lots of examples,  it has never been proven rigorously under Assumptions 1 and 2.
In section 2, we prove Theorem \ref{MT}.  In  section 3, 
we apply Theorem \ref{MT} to  spontaneous symmetry-breaking phenomena
in some examples. Theorem \ref{MT} indicates that replica symmetry-breaking  in disordered spin systems should be a spontaneous symmetry-breaking.

\section{Proof}
To prove the following lemmas, here, we define the Duhamel product. 
Define a fictitious time  $t \in [0,1]$ and a time evolution of operators with the Hamiltonian.
Let $O$ be an arbitrary self-adjoint operator, and we define an operator valued function  $O(t)$ of $t\in[0,1]$  by
\begin{equation}
 O(t):= e^{-tH} O  e^{tH}.
\end{equation}
The  Duhamel product of time independent operators $O_1,O_2, \cdots, O_k$ is defined
by
 \begin{equation}
( O_1,O_2, \cdots, O_k)_{{\bf x}} :=\int_{[0, 1]^k} dt_1\cdots dt_k \langle {\rm T}[ O_1(t_1)  O_2(t_2) \cdots  O_k(t_k) ]\rangle_N,_{{\bf x}},
 \end{equation}
where the symbol ${\rm T}$ is a multilinear mapping of the chronological ordering.
If we define a partition function with arbitrary self-adjoint operators  $O_1, \cdots, O_k$ and real
numbers $x_1, \cdots, x_k$
 \begin{equation}
Z(x_1,\cdots, x_k) := {\rm Tr} \exp \beta \left[-H+\sum_{i=1} ^k x_i O_i \right],
 \end{equation}
the Duhamel product of $k$ operators represents
 the $k$-th order derivative of the partition function 
 \cite{Cr,GUW,S}
 \begin{equation}\beta^k( O_1,\cdots, O_k)_{{\lambda}}=\frac{1}{Z({\bf x})}
\frac{\partial ^k Z({\bf x})}{\partial x_1 \cdots \partial  x_k}.
 \end{equation}
Furthermore,  a truncated Duhamel product is defined by 
 \begin{equation}\beta^k( O_1 ; \cdots ;   O_k)_{\bf x}=
\frac{\partial ^k }{\partial x_1 \cdots \partial  x_k} \log Z({\bf x}).
 \end{equation}
 Note that the Duhamel product of a single operator is identical to its Gibbs expectation
\begin{equation}\beta \langle O_1 \rangle_N,_{\bf x}=\beta( O_1)_{\bf x}=\frac{1}{Z({\bf x})}
\frac{\partial  Z({\bf x})}{\partial x_1}=
\frac{\partial }{\partial x_1} \log Z({\bf x}) .
 \end{equation}

 {\lemma \label{varpsi} 
There exists a positive number $K$ independent of the system size $N$, such that  the variance of $\psi_N$ is bounded by 
\begin{equation}
{\mathbb E} \psi_N(\beta,\lambda, {\bf J}) ^2- p_N(\beta,\lambda)^2 \leq \frac{K}{N},
 \end{equation}
 for any $(\beta, \lambda)\in (0,\infty) \times {\mathbb R}.$\\

\noindent Proof.}  
Fix a  bijection
$\cup_{p=x,y,z} {\cal C}_N^p \to [1,M] \cap {\mathbb Z}$ arbitrarily for identification  $\cup_{p=x,y,z} {\cal C}_N^p = [1,M] \cap {\mathbb Z}$ for $M:=|\cup_{p=x,y,z} {\cal C}_N^p|$, such that
this bijection numbers the sequence ${\bf J}$ as $(J_X^p)_{X\in {\cal C}_N^p, p=x,y,z}  = (J_j)_{j=1,2,\cdots, M}$. 

For an integer  $1 \leq m \leq M$,  define a symbol 
${\mathbb E}_m$ which denotes the expectation over 
random variables  $( J_j)_{j > m}$. Note that ${\mathbb E}_0 ={\mathbb E}$ is the expectation over  all random variables $(J_j)_{j=1,2, \cdots, M} $,
and ${\mathbb E}_M$ is the identity.
 Here, we represent $\psi_N({\bf J})$ as a function of a sequence of random variables ${\bf J}=(J_j)_{j =1, \cdots, M}$ for lighter notation. 
 \begin{eqnarray}
&& {\mathbb E} \psi_N({\bf J})^2- ({\mathbb E} \psi_N({\bf J}))^2\\
&&= {\mathbb E} ({\mathbb E}_M \psi_N({\bf J}) )^2
- {\mathbb E} ({\mathbb E}_0 \psi_N({\bf J}))^2\\
&&=\sum_{m=1} ^M{\mathbb E}[ ({\mathbb E}_m \psi_N( {\bf J}) )^2
- ({\mathbb E}_{m-1} \psi_N( {\bf J}))^2].
 \end{eqnarray}
 In the $m$-th term, regard $\psi_N (J_m)$ as a function of $J_m$ for simplicity.
Let $J_m'$ be an independent random variable satisfying the same distribution as that of $J_m$, and 
${\mathbb E}'$ denotes an expectation over only $J'_m$.  In this notation, 
\begin{eqnarray}
&&{\mathbb E}_{m-1} \psi_N(J_m) = {\mathbb E}_{m-1} \psi_N({\bf J})  \nonumber \\ 
&&={\mathbb E}_{m-1} \psi_N(J_1, \cdots,J_{m-1}, J_m, J_{m+1},\cdots, J_M) \nonumber \\
&&={\mathbb E}_{m}{\mathbb E}' \psi_N(J_1, \cdots, J_{m-1} ,J_m', J_{m+1}, \cdots, J_M) \nonumber \\
&&={\mathbb E}_{m}{\mathbb E}' \psi_N(J_m').\nonumber
\end{eqnarray}
Therefore, a bound on the $m$-th term is given by
\begin{eqnarray}
&&{\mathbb E}[ ({\mathbb E}_m \psi_N(J_m ))^2
- ({\mathbb E}_{m-1} \psi_L(J_m))^2] \\
&&
={\mathbb E} [({\mathbb E}_m \psi_N( J_m ))^2
- ({\mathbb E}_m{\mathbb E}' \psi_N( J'_m))^2] \\&&
=  {\mathbb E}
[ {\mathbb E}_m( \psi_N(J_m) -  {{\mathbb E}}' \psi_N( J_m') ) ]^2\\
&&={\mathbb E} [ {\mathbb E}_m {{\mathbb E}}'( \psi_N( J_m ) -   \psi_N( J_m')) ]^2\\
&&
={\mathbb E} \Big[{\mathbb E}_m  {{\mathbb E}}' \int_{J_m'} ^{J_m} dJ \frac{\partial }{\partial J} \psi_N( J)\Big]^2,\\
&&
=\frac{1}{N^2}{\mathbb E} \Big[{\mathbb E}_m  {{\mathbb E}}' \int_{J_m'} ^{J_m} dJ \beta \langle \varphi_m(S_m) \rangle_N,_{ J} \Big]^2\\
&&
=\frac{1}{N^2}{\mathbb E}  \Big[ {\mathbb E}_m  {{\mathbb E}}' \frac{J_m-J_m'}{J_m-J_m'} \int_{J_m'} ^{J_m} dJ \beta \langle \varphi_m(S_m) \rangle_N,_{ J} \Big]^2\\
&& 
\leq\frac{1}{N^2}{\mathbb E} {{\mathbb E}}'  \Big[ \frac{J_m-J_m'}{J_m-J_m'} \int_{J_m'} ^{J_m} dJ \beta \langle \varphi_m(S_m) \rangle_N,_{ J} \Big]^2\\
&& 
\leq\frac{1}{N^2}{\mathbb E} {{\mathbb E}}'  \frac{(J_m-J_m')^2}{J_m-J_m'} \int_{J_m'} ^{J_m} dJ\Big[  \beta \langle \varphi_m(S_m) \rangle_N,_{ J} \Big]^2\\
&& 
\leq \frac{\beta^2 C_\varphi^2}{N^2}{\mathbb E}  {\mathbb E}'  (J_{m}-J_{m}')^2 = \frac{2\beta^2C_\varphi^2\sigma^2 }{N^2},
 \end{eqnarray}
 where we have used  Jensen's inequality. Note that 
  there exist $p=x,y,z$ and  $X \in {\cal C}_N^p$ and $S_m = S_X^p$. 
 Therefore 
\begin{equation}
{\mathbb E} \psi_N({\bf J})^2- ({\mathbb E} \psi_N( {\bf J}))^2 \leq \sum_{m=1}^M
 \frac{2 \beta^2 C_\varphi^2\sigma^2}{N^2}
 \leq \frac{2\beta^2C_\varphi^2\sigma^2}{N}.
\end{equation}
This completes the proof, if  $K$ denotes $2\beta^2C_\varphi^2\sigma^2$. $\Box$\\

The following lemma can be shown by the standard convexity argument to imply the Ghirlanda-Guerra identities  \cite{AC,C2,CG2,GG,I,I2,Pn,T}
in classical and quantum disordered systems. 
The proof can be done on the basis of of convexity  of functions $\psi_N$, $p_N$, $p$ and their almost everywhere differentiability and 
Assumptions 1 and 2.

{\lemma \label{Delta} For  almost all $\lambda\in {\mathbb R}$, 
the infinite-volume limit of the expectation of ${\cal O}_N$
\begin{equation}
 \lim_{N\to\infty}  {\mathbb E} \langle {\cal O}_N \rangle_N,_{\lambda} =\frac{1}{\beta} \frac{\partial p}{\partial \lambda} (\beta, \lambda) 
\label{lim1}
\end{equation}
exists  and the following variance vanishes
\begin{equation}
\lim_{N\to\infty} [{\mathbb E} \langle {\cal O}_N \rangle_N,_{\lambda}^2 -({\mathbb E} \langle {\cal O}_N \rangle_N,_{\lambda}) ^2]=0,
\end{equation}
for each $\beta \in (0,\infty)$.\\

\noindent
Proof.}
First, let us regard $p_N({\lambda})$, $p({\lambda})$ and $\psi_N({\lambda})$ as functions of $\lambda$ for lighter notation. 
Define the following functions of $\epsilon > 0$.
\begin{eqnarray}
&&w_N(\epsilon) := \frac{1}{\epsilon}[|\psi_N(\lambda+\epsilon )-p_N({\lambda}+\epsilon)|+|\psi_N({\lambda}- \epsilon)-p_N({\lambda}-\epsilon)|
+|\psi_N({\lambda} )-p_N({\lambda})| ], \nonumber 
\\
&&e_N(\epsilon ):=\frac{1}{\epsilon}[|p_N(\lambda+\epsilon )-p({\lambda}+\epsilon)|+|p_N({\lambda}- \epsilon)-p({\lambda}-\epsilon)|
+|p_N({\lambda} )-p({\lambda})|].
\end{eqnarray}
Assumption 1  and Lemma \ref{varpsi} give
\begin{equation}
\lim_{N\to\infty} {\mathbb E}w_N(\epsilon)=0, \ \ \  \lim_{N\to\infty} e_N(\epsilon)=0,
\end{equation}
for any $\epsilon > 0$. 
Since $\psi_N$, $p_N$ and $p$ are convex functions of ${\lambda}$, we have
\begin{eqnarray} 
&&\frac{\partial \psi_N}{\partial \lambda}(\lambda) - \frac{\partial  p}{\partial {\lambda}}({\lambda})
\leq \frac{1}{\epsilon} [\psi_N({\lambda}+\epsilon)-\psi_N({\lambda})]- \frac{\partial  p}{\partial {\lambda}}
\\
&&\leq \frac{1}{\epsilon} [\psi_N({\lambda}+\epsilon)-p_N({\lambda}+\epsilon)+p_N({\lambda}+\epsilon)-p_N({\lambda})
+p_N({\lambda})-\psi_N({\lambda}) \nonumber \\
&& -p({\lambda}+\epsilon) +p({\lambda}+\epsilon)+p({\lambda})-p({\lambda}) ]- \frac{\partial  p}{\partial {\lambda}}({\lambda}) 
\\
&&\leq \frac{1}{\epsilon} [ |\psi_N({\lambda}+\epsilon)-p_N({\lambda}+\epsilon)|
+|p_N({\lambda})-\psi_N({\lambda})| +|p_N({\lambda}+\epsilon)-p({\lambda}+\epsilon)|\nonumber 
\\ 
&&+|p_N({\lambda})-p({\lambda})| ]+\frac{1}{\epsilon}[ p({\lambda}+\epsilon)-p({\lambda})] - \frac{\partial  p}{\partial {\lambda}}({\lambda}) 
 \\
&&\leq w_N(\epsilon) +e_N(\epsilon)
+  \frac{\partial p}{\partial {\lambda}}({\lambda}+\epsilon) - \frac{\partial  p}{\partial {\lambda}}({\lambda}). 
\end{eqnarray}  
As in the same calculation, we have
\begin{eqnarray} 
&&\frac{\partial \psi_N}{\partial {\lambda}}({\lambda}) - \frac{\partial  p}{\partial {\lambda}}({\lambda}) 
\geq \frac{1}{\epsilon}[\psi_N({\lambda})-\psi_N({\lambda}-\epsilon)] - \frac{\partial  p}{\partial {\lambda}}({\lambda}) 
\\&&
\geq -w_N(\epsilon) -e_N(\epsilon)+ \frac{\partial p}{\partial {\lambda}}({\lambda}-\epsilon)- \frac{\partial  p}{\partial {\lambda}}({\lambda}) . 
\end{eqnarray}  
Then, 
\begin{eqnarray} 
{\mathbb E}\Big|\frac{\partial \psi_N}{\partial {\lambda}}({\lambda}) - \frac{\partial p}{\partial {\lambda}}({\lambda})\Big| \leq {\mathbb E}w_N(\epsilon)
+e_N(\epsilon)+  \frac{\partial p}{\partial {\lambda}}({\lambda}+\epsilon) -  \frac{\partial p}{\partial {\lambda}}({\lambda}-\epsilon).
\end{eqnarray}  
Convergence of $p_N$ in  the infinite-volume limit implies 
\begin{eqnarray}
&&\lim_{N\to\infty } {\mathbb E}\Big| \beta \langle {\cal O}_N \rangle_N,_\lambda - \frac{\partial p}{\partial {\lambda}}({\lambda})\Big|
\leq  \frac{\partial p}{\partial {\lambda}}({\lambda}+\epsilon) -  \frac{\partial p}{\partial {\lambda}}({\lambda}-\epsilon),
\end{eqnarray}
The right-hand side vanishes, since the convex function $p({\lambda})$ is continuously 
differentiable almost everywhere and $\epsilon >0$ is arbitrary. Therefore
\begin{equation}
\lim_{N\to\infty}{\mathbb E} \Big| \beta  \langle {\cal O}_N \rangle_N,_\lambda - \frac{\partial p}{\partial {\lambda}}({\lambda})\Big|=0.
\label{limit3}
\end{equation}
for almost all ${\lambda}$.   Jensen's inequality gives 
\begin{equation}
\lim_{N\to\infty}\Big| {\mathbb E}\beta  \langle {\cal O}_N \rangle_N,_\lambda - \frac{\partial p}{\partial {\lambda}}( \lambda)\Big|=0.
\end{equation}
 This implies the first equality (\ref{lim1}).  
Since the  $p({\lambda})$ is continuously differentiable almost everywhere in ${\mathbb R}$, 
these equalities imply also
 \begin{equation}
\lim_{N\to\infty }  {\mathbb E} |\langle {\cal O}_N \rangle_N,_\lambda - 
 {\mathbb E} \langle {\cal O}_N  \rangle_N,_\lambda| =0.
 \end{equation}
The bound on ${\cal O}_N$  implies the following limit
 \begin{equation}
\lim_{N\to\infty }  {\mathbb E} (\langle {\cal O}_N \rangle_N,_\lambda - 
 {\mathbb E} \langle {\cal O}_N  \rangle_N,_\lambda)^2 \leq 
 2C_o \lim_{N\to\infty }  {\mathbb E} |\langle {\cal O}_N \rangle_N,_\lambda - 
 {\mathbb E} \langle {\cal O}_N  \rangle_N,_\lambda|=0.
 \end{equation}
This  completes the proof.
$\Box$\\

Note that Lemma \ref{Delta} guarantees the existence of the following 
infinite-volume limit for almost all $\lambda\in {\mathbb R}$
 \begin{equation}
\lim_{N\to\infty}{ \mathbb E}\langle {\cal O}_N \rangle_N,_\lambda^2 
=(\lim_{N\to\infty}{ \mathbb E}\langle {\cal O}_N \rangle_N,_\lambda)^2
 \end{equation} 

{\lemma 
 \label{delta} For  almost all $\lambda\in {\mathbb R}$, 
  the  following variance  of ${\cal O}_N$ vanishes in the infinite-volume limit
\begin{equation}
\lim_{N\to \infty} {\mathbb E} \langle ({\cal O}_N,_\lambda  -\langle {\cal O}_N \rangle_N,_\lambda  )^2 \rangle_N,_\lambda = 0.
\end{equation}
Proof.} The derivative of the Gibbs expectation is represented in the Duhamel product
\begin{equation}
\frac{\partial}{\partial {\lambda}}  {\mathbb E} \langle {\cal O}_N \rangle_N, =N \beta  {\mathbb E} [ ({\cal O}_N, {\cal O}_N)_{N,{\lambda}} -\langle {\cal O}_N\rangle_N,_{{\lambda}}^2 ],
\label{dh}
\end{equation}
The integration of the both sides in (\ref{dh})  over an arbitrary interval $ (\lambda',\lambda'')$  with $\lambda' < \lambda''$ gives
\begin{equation}
 {\mathbb E} \langle {\cal O}_N \rangle_N,_{\lambda''} - {\mathbb E} \langle {\cal O}_N\rangle_N,_{\lambda'} =N \beta \int_{\lambda'} ^{\lambda''} d{\lambda}   {\mathbb E}   [ ({\cal O}_N, {\cal O}_N)_{N,\lambda} -\langle {\cal O}_N\rangle_N,_{\lambda}^2 ].
\end{equation}
Since ${\cal O}_N $ has a uniform bound, we have 
 \begin{equation}
  \int_{\lambda'} ^{\lambda''} d{\lambda}  {\mathbb E} [ ({\cal O}_N, {\cal O}_N)_{\lambda} -\langle {\cal O}_N\rangle_N,_{\lambda}^2 ] 
= \frac{1}{N\beta} {\mathbb E}  (\langle {\cal O}_N \rangle_N,_{\lambda''} -\langle {\cal O}_N\rangle_N,_{\lambda'}) \leq \frac{2C_o}{N\beta}.
\end{equation}
The positive semi-definiteness of the integrand gives
 \begin{equation}
\lim_{N\to\infty}  \int_{\lambda'} ^{\lambda''} d{\lambda}  {\mathbb E} [ ({\cal O}_N, {\cal O}_N)_{N,\lambda} -\langle {\cal O}_N\rangle_N,_{\lambda}^2 ] 
= 0.
\end{equation}
The integrand has uniform bounds
\begin{equation}
0 \leq {\mathbb E} [ ({\cal O}_N, {\cal O}_N)_{N,\lambda} -\langle {\cal O}_N\rangle_N,_{\lambda}^2 ] \leq {\mathbb E}  \langle {\cal O}_N^2\rangle_N,_{\lambda}  
\leq C_o^2,
\end{equation}
thus Lebesgue's dominated convergence theorem gives
 \begin{equation}
 \int_{\lambda'} ^{\lambda''} d{\lambda} \limsup_{N\to\infty}   {\mathbb E} [ ({\cal O}_N, {\cal O}_N)_{N,\lambda} -\langle {\cal O}_N\rangle_N,_{\lambda}^2 ] 
=\lim_{N\to\infty}  \int_{\lambda'} ^{\lambda''} d{\lambda}  {\mathbb E} [ ({\cal O}_N, {\cal O}_N)_{N,\lambda} -\langle {\cal O}_N\rangle_N,_{\lambda}^2 ] 
= 0.
\end{equation}
Since the integration  interval $ (\lambda',\lambda'')$ is arbitrary, we have 
\begin{equation}
\lim_{N\to\infty}  {\mathbb E}   [ ({\cal O}_N, {\cal O}_N)_{N,\lambda} -\langle {\cal O}_N\rangle_N,_{\lambda}^2 ] =0,
\label{limd}
\end{equation}
for almost all $\lambda \in {\mathbb R}$.
Harris'  inequality of the Bogolyubov type
between the Duhamel function and the Gibbs expectation of the square of arbitrary self-adjoint operator ${\cal O}_N$ \cite{BT,H}
\begin{equation}
 \langle {\cal O}_N^2\rangle_N,_{\lambda} - \frac{\beta}{12} \langle[ {\cal O}_N, [H_{\lambda} ,{\cal O}_N]] \rangle_N,_{\lambda}\leq ( {\cal O}_N,  {\cal O}_N)_{N,\lambda}  \leq \langle {\cal O}_N^2\rangle_N,_{\lambda}.
\label{harris}
\end{equation}
These inequalities, Assumption 2 and the limit (\ref{limd})  imply
\begin{equation}
\lim_{N\to\infty}  {\mathbb E}   \langle ( {\cal O}_N -\langle {\cal O}_N\rangle_N,_{\lambda} )^2 \rangle_N,_{\lambda}
=\lim_{N\to\infty}  {\mathbb E}  (\langle {\cal O}_N^2\rangle_N,_{\lambda} -\langle {\cal O}_N\rangle_N,_{\lambda}^2 )=0.
\label{lims}
\end{equation}
This  completes the proof. $\Box$

\paragraph{Proof of Theorem \ref{MT}}
Lemma \ref{Delta} and Lemma \ref{delta} imply
\begin{equation}
\lim_{N\to\infty} {\mathbb E} [\langle {\cal O}_N ^2\rangle_N,_\lambda - ({\mathbb E} \langle {\cal O}_N \rangle_N,_\lambda )^2 ] =
 \lim_{N\to\infty} {\mathbb E} [\langle {\cal O}_N^2\rangle_N,_\lambda-{\mathbb E} \langle {\cal O}_N \rangle_N,_\lambda^2 +{\mathbb E} \langle {\cal O}_N \rangle_N,_\lambda^2
  - ({\mathbb E} \langle {\cal O}_N \rangle_N,_\lambda )^2 ] =0.
\end{equation}
This completes the proof. $\Box$
\section{Applications}

\subsection{Spontaneous symmetry-breaking in quantum spin systems}
First, we remark the general properties of spontaneous symmetry-breaking in quantum spin  systems without disorder.
Spontaneous symmetry-breaking phenomena are observed generally in many-body systems possessing some symmetries.    
It is well-known that the quantum Heisenberg model gives a  simple example of spontaneous symmetry-breaking \cite{BR}. 
The SU(2) 
symmetry-breaking occurs in the  ferromagnetic or antiferromagnetic phases in the Heisenberg model. 
First, we  give a brief review of the relation between the long-range order and the spontaneous symmetry-breaking.
Then, we describe the property of the symmetric and  the symmetry-breaking Gibbs states
from the  point of view of Theorem \ref{MT} and the law of large numbers which plays a fundamental role in statistical physics. 
Define a Hamiltonian $H_N({\bf S})$ of the Heisenberg model 
 with a sequence of coupling constants  ${\bf J} = (J_{i,j})_{\{i,j\} \in B_N; p=x,y,z}$ 
\begin{equation}
 H_N({\bf S}, {\bf J}) := \sum_{\{i,j\} \in B_N} J_{i,j}\sum_{p =x,y,z}  S^p_i S^p_j,
\label{hamilhei}
\end{equation}
where $B_N$ is a set of bonds  which consists of $i,j \in V_N$ 
under a certain condition, such as nearest-neighbor interaction $|i-j|=1$.  
The Hamiltonian is invariant under an arbitrary SU(2) transformation $U = \bigotimes_{j\in V_N}  U_j$
$$
H_N( U {\bf S} U^{-1},{\bf J} ) = H_N({\bf  S},{\bf J}),
$$
where   $U_j$ is in SU(2) acting on ${\cal H}_j$.
The partition function is defined on the basis of this Hamiltonian
\begin{equation}
Z_{N}(\beta, {\bf J}) := {\rm Tr} e^{ - \beta H({\bf S}, {\bf J})   }.
\end{equation}
 The  SU(2) symmetric  Gibbs expectation for an arbitrary function $f({\bf S})$ of spin operators  is
\begin{equation}
\langle f({\bf S}) \rangle_N, = \frac{1}{Z_N(\beta,{\bf J})} {\rm Tr} f({\bf S} ) e^{-\beta H({\bf S},{\bf J} )}.  
\end{equation} 
  To detect the symmetry-breaking, define an order operator with a sequence $(a_j)_{j\in V_N} \in {\mathbb R}$ for $p=x,y,z$
\begin{equation}
{\cal O}_N := \frac{1}{N} \sum_{j\in V_N} a_j S_j^p.
\end{equation}
For example,  the sequence $(a_j)_{j\in V_N}$ is defined by
 $a_j =1$  for the ferromagnetic order , and  by $a_j =(-1)^{j_1+ \cdots + j_d} $ for antiferromagnetic order  at $j= (j_1, \cdots, j_d) \in V_N$ 
in $d$ dimensional cubic lattice $V_N =[1,L]^d\cap{\mathbb Z}^d$.
The order operator transforms $U {\cal O}_NU^{-1} \neq {\cal O}_N$ under an arbitrary SU(2) transformation $U$ and it has a uniform bound
$\| {\cal O}_N \| \leq C_o$ with a positive constant independent of the system size $N$. 
Then, the SU(2) symmetric Gibbs expectation of the order operator vanishes 
\begin{equation}
\langle {\cal O}_N \rangle_N, =0.
\end{equation} 
Dyson, Lieb and Simon proved  that the symmetric Gibbs expectation of the square of the order operator
does not vanishes for $d\geq 3 $ with the antiferromagnetic  nearest-neighbor interactions and for sufficiently large $\beta>0$
$$
\lim_{N\to\infty}\langle {\cal O}_N^2\rangle_N, >0,
$$
in the infinite-volume limit \cite{DLS}.
This phenomenon is long-range order. For the SU(2) invariant ferromagnetic interaction, the existence of the long-range order is believed, 
nonetheless it has never been proven at finite temperature.   The long-range order and the 
vanishing Gibbs expectation of the order operator  imply
that the variance of ${\cal O}_N$ in the symmetric Gibbs state does not vanish in the infinite-volume limit.
It is believed that  the expectation value calculated in the Gibbs state  can be identical to its observed value, if 
its variance vanishes.  This is guaranteed by the law of large numbers, which is proven by the Chebyshev inequality. 
If the variance becomes finite, the Paley-Zygmund  inequality  rules out its identification between the expectation value and the observed value.
It is believed that  the symmetry-breaking occurs instead of 
the violation of the law of  large numbers.  To consider the spontaneous symmetry-breaking, apply a symmetry-breaking perturbation
in the Hamiltonian
\begin{equation}
H_\lambda:= H({\bf S}, {\bf J} ) - N \lambda {\cal O}_N,
\end{equation} 
and define an expectation of the function $f({\bf S})$ 
in a symmetry-breaking Gibbs state 
\begin{equation}
\langle f({\bf S}) \rangle_N,_\lambda :=  \frac{1}{Z_N(\beta, \lambda, {\bf J})} {\rm Tr} f({\bf S}) e^{-\beta H_\lambda},
\end{equation}
where the partition function is 
$$
Z_N(\beta, \lambda, {\bf J}) := {\rm Tr} f({\bf S}) e^{-\beta H_\lambda }.
$$
If spontaneous symmetry-breaking occurs, the following expectation of an order operator
\begin{equation}
\lim_{\lambda \searrow 0} \lim_{N\to \infty} 
\langle {\cal O}_N \rangle_N,_\lambda \neq 0,
\end{equation} 
exists as a non-zero value in the infinite-volume and switched-off limits.
In this case,  the two limiting procedures  
do not commute
\begin{equation}
0\neq \lim_{\lambda \searrow 0 } \lim_{N\to \infty} 
\langle {\cal O}_N \rangle_N,_\lambda  \neq \lim_{N\to \infty} \lim_{\lambda \searrow 0}  
\langle {\cal O}_N \rangle_N,_\lambda =0.
\end{equation}
The Griffiths-Koma-Tasaki theorem indicates the inequality between the long-range order and the spontaneous symmetry-breaking \cite{Gff,KT}.
The standard deviation of  an order operator ${\cal O}_N$ in the symmetric Gibbs state
is bounded by its expectation in the  switched-off limit of the symmetry-breaking perturbation after the infinite-volume limit.   
\begin{equation}
\lim_{N\to \infty} 
\sqrt{ \langle {\cal O}_N^2 \rangle_N,_0 } \leq  \lim_{\lambda \searrow 0} \lim_{N\to \infty} 
\langle {\cal O}_N \rangle_N,_\lambda.    
\end{equation}
The existence of the long-range order in the symmetric Gibbs state implies the existence of corresponding  spontaneous symmetry-breaking. 
Theorem \ref{MT} shows that  the variance of the order operator vanishes 
always in the symmetry-breaking Gibbs state
\begin{equation}
\lim_{N\to\infty} [\langle {\cal O}_N^2 \rangle_N,_\lambda  -\langle {\cal O}_N \rangle_N,_\lambda^2  ]=0,
\label{vv}
\end{equation}
since the concerned model satisfies Assumptions 1 and 2. Theorem \ref{MT} and the existence of long-range order in the symmetric Gibbs state
 imply the non-commutativity between the infinite-volume limit  $N\to\infty$ and switched-off  limit  $\lambda \searrow  0$,
since the above infinite-volume limit  is valid also in the switched  off limit  of the uniform field $\lambda \searrow  0$ or $\lambda \nearrow  0$.
In addition to this fact,
 the expectation  $\langle {\cal O}_N\rangle_N,_{\pm0}$
in the symmetry-breaking  
Gibbs state should be identical to the corresponding observed value of $\cal O$ according to the law of large numberss.   The symmetry-breaking should  occur instead of 
the violation of the law of  large numberss, because of the  limit (\ref{vv}).

\subsection{Replica symmetry-breaking in disordered spin systems}

Replica symmetry-breaking (RSB)  phenomena have been studied  in extensive fields in science
in addition to statistical physics, since Parisi found the replica symmetry-breaking formula for the Sherrington-Kirkpatrick (SK) model \cite{SK} which gave a great  breakthrough
in the theory of disordered spin systems \cite{Pr}.   In low-temperature phase of the SK model, the distribution of the spin overlap becomes broadened, which shows  RSB.
Since Talagrand  proved the Parisi formula rigorously \cite{T2,T},  mathematicians have been studying RSB phenomena in many mathematical systems.
Recently, Chatterjee  has proven that there is no RSB phase in the random field Ising model \cite{C2}.  He defines RSB in terms of the replica symmetric Gibbs state and the sample distribution, then he has proven that the variance of the spin overlap  vanishes in the replica symmetric Gibbs state and the sample expectation. 
Note that Chattejee's definition does not exclude a possibility of RSB  without spontaneous symmetry-breaking. 
Here,  we give an extension of Chatterjee's definition of RSB and  
prove that  RSB should occur as a spontaneous symmetry-breaking  phenomenon in disordered quantum spin systems.   
To study spontaneous RSB, we apply a RSB perturbation to the system, as discussed in the random energy model \cite{G2,M2}.
Then, we prove that the  commutativity  between the infinite-volume limit and 
the  replica symmetric limit implies the absence of RSB in Chatterjee's definition.

Consider $n$ replicated spin operators $(S^{p,\alpha}_i)_{i \in V_N;  p=x,y,z; \alpha=1, \cdots, n}$ and replica symmetric  Hamiltonian
\begin{equation}
H({\bf S}^1, \cdots, {\bf S}^n, {\bf J}) :=\sum_{\alpha=1}^n  H_N({\bf S}^\alpha, {\bf J}).
\label{replicahamil}
\end{equation}
The replica symmetry is a permutation symmetry among these replicated spins.
The replica symmetry is the invariance of the Hamiltonian under an arbitrary permutation $\sigma$ on $\{1, \cdots, n\}$   
$$
H({\bf S}^1, \cdots, {\bf S}^n, {\bf J})=H({\bf S}^{\sigma 1}, \cdots, {\bf S}^{\sigma n}, {\bf J})
$$
Define a spin overlap by
\begin{equation}
R_{\alpha,\beta}^p := \frac{1}{|{\cal D}_N^p |}\sum_{X \in {\cal D}_N^p} S_X^{p,\alpha} S_X^{p,\beta}.
\end{equation}
where ${\cal D}_N^p$ is a certain collection of subsets $X \subset  V_N$. 
Note that the replica symmetric Gibbs expectation of the overlap becomes
 \begin{equation}
\langle R_{\alpha,\beta}^p\rangle_N,  = \frac{1}{|{\cal D}_N^p |}\sum_{X \in {\cal D}_N^p}\langle  S_X^{p,\alpha} S_X^{p,\beta} \rangle_N,
= \frac{1}{|{\cal D}_N^p |}\sum_{X \in {\cal D}_N^p}\langle  S_X^p\rangle_N,^2.
\end{equation}
For ${\cal D}_N^p=V_N$ with $p=z$ in the Ising model, this Gibbs expectation
 is the Edwards-Anderson spin glass order parameter \cite{EA}, as mathematically studied by
van Enter and Griffiths \cite{VG}.
Define a replica symmetric Gibbs state with the Hamiltonian (\ref{replicahamil}).
For  $\beta>0$,  the  partition function is defined by
\begin{equation}
Z_{N,n}(\beta, {\bf J}) := {\rm Tr} e^{ - \beta H({\bf S}^1, \cdots, {\bf S}^n, {\bf J})   }
\end{equation}
where the trace is taken over the Hilbert space $\bigotimes_{\alpha=1}^n {\cal H}$.
Let  $f$ be an arbitrary function 
of  $n$ replicated  spin operators.  
The  expectation of $f$ in the Gibbs state is given by
\begin{equation}
\langle f({\bf S}^1, \cdots, {\bf S}^n) \rangle_N,=\frac{1}{Z_{N,n}(\beta, {\bf J})}{\rm Tr} f({\bf S}^1, \cdots, {\bf S}^n)  e^{ - \beta H({\bf S}^1, \cdots, {\bf S}^n, {\bf J})  }.
\end{equation}
Let us define RSB  in an extension of Chatterjee's definition \cite{C2}. For simplicity, we assume that
 the Gibbs state has only replica symmetry. If the Hamiltonian has other symmetry, we remove it from the Gibbs state by
 some  suitable perturbation or boundary condition.  
{\definition \label{cRSB}
Consider  the replica symmetric Gibbs state of a disordered quantum spin model defined by the replica symmetric Hamiltonian
(\ref{replicahamil}) and assume that the Gibbs state has no other symmetry.  
We say that  replica symmetry-breaking (RSB)  does not occur in Chatterjee's sense, 
if the variance of  any  RSB order operator ${\cal R}$ 
whose expectation value ${\mathbb E} \langle {\cal R} \rangle_N,$ exists in the infinite-volume limit, 
vanishes in the infinite-volume limit
\begin{equation}
\lim_{N \to \infty }{\mathbb E}\langle ( {\cal R}  - {\mathbb E} \langle {\cal R}\rangle_N,   )^2\rangle_N,  = 0.
\end{equation} 
}
 This variance is decomposed into the following  two terms
\begin{equation}
{\mathbb E}\langle ( {\cal R} -{\mathbb E} \langle {\cal R}  \rangle_N, )^2 \rangle_N,
= {\mathbb E}\langle ( {\cal R}   - \langle {\cal R}  \rangle_N,)^2 \rangle_N, +
{\mathbb E} ( \langle  {\cal R} \rangle_N,  -  {\mathbb E} \langle {\cal R}  \rangle_N,)^2. 
 \end{equation}
 The finiteness of the first term in the right-hand side
 implies that the observed value of RSB order operator differs from its replica
  symmetric  Gibbs state in some samples, then this means spontaneous RSB. 
  This imples that there are several samples where the observed spin  overlap is not identical to its Gibbs expectation. 
 This seems like the long-range order in the SU(2) invariant Gibbs state in the Heisenberg model, when spontaneous SU(2) symmetry-breaking occurs 
 \cite{Gff,KT}. On the other hand, the finiteness of the second term implies that  the replica symmetric Gibbs expectation of the RSB order operator  in an 
 arbitrary  sample differs from its sample expectation.
 In classical Ising systems with Gaussian disorder and with
 ${\cal D}_N^p={\cal C}_N^p$, such as the SK model, 
 the Ghirlanda-Guerra identities \cite{AC,C2,C1,CG,CG2,GG} give the relation between two terms \cite{C2}
 \begin{equation}
 {\mathbb E}\langle ( R_{1,2}^z    - \langle R_{1,2}^z  \rangle_N,)^2 \rangle_N, =\frac{2}{3}
{\mathbb E} \langle (   R_{1,2}^z  -  {\mathbb E} \langle R_{1,2}^z  \rangle_N,)^2 \rangle_N,. 
 \end{equation}
 This identity implies that a non-zero value of the right-hand side is equivalent to a non-zero value of the left hand side.
Therefore, replica symmetry-breaking occurs  always as a spontaneous symmetry-breaking
in this case. Next, we consider general cases, for example ${\cal D}_N^p \neq {\cal C}_N^p$ or disordered quantum spin systems
on the basis of Theorem \ref{MT}.\\

To study spontaneous replica symmetry-breaking, we apply a RSB order operator as a perturbation to the replica symmetric Hamiltonian.
Let  ${\cal R}$ be  a  RSB perturbation operator which is  self-adjoint operator with a uniform  bound $\| {\cal R}\| \leq C_R$. 
Define a perturbed  Hamiltonian by
\begin{equation}
H_\lambda ({\bf S}_1 , \cdots,  {\bf S}_n, {\bf J} ) = H({\bf S}^1, \cdots, {\bf S}^n, {\bf J}) - N \lambda {\cal R} ,
\label{RSBhamil}
\end{equation} 
with   coupling constants $\lambda \in {\mathbb R}$.
For example,   ${\cal R}$ is given by a linear combination of functions of spin overlaps 
\begin{equation}\displaystyle {\cal R} =\sum_{a \in A} c_a  (R_{1,2}^p)^a.
\label{calR}
\end{equation}
Define the Gibbs state with the Hamiltonian (\ref{RSBhamil}).
For  $\beta>0$,  the  partition function is defined by
\begin{equation}
Z_{N,n}(\beta,\lambda, {\bf J}) := {\rm Tr} e^{ - \beta  H_\lambda ({\bf S}_1 , \cdots,  {\bf S}_n, {\bf J} ) }.
\end{equation}
Let  $f$ be an arbitrary function 
of $n$ replicated spin operators.  
The  expectation of $f$ in the Gibbs state is given by
\begin{equation}
\langle f({\bf S}^1, \cdots, {\bf S}^n) \rangle_N,_\lambda=\frac{1}{Z_N(\beta, \lambda, {\bf J})}{\rm Tr}
 f({\bf S}^1, \cdots, {\bf S}^n)  e^{ - \beta  H_\lambda({\bf S}_1 , \cdots,  {\bf S}_n, {\bf J} )}.
\end{equation}
Define the following function from the partition function
$$
\psi_{N,n}(\beta, \lambda,{\bf J}):=\frac{1}{N} \log Z_{N,n}(\beta,\lambda, {\bf J}).
$$
and its expectation $$p_{N,n}(\beta, \lambda):={\mathbb E} \psi_{N,n}(\beta, \lambda, {\bf J}).$$

Next we remark a property of  spontaneous replica symmetry-breaking.  
{\note  \label{sRSB} Consider a RSB Gibbs state of a disordered quantum spin model defined by the Hamiltonian
(\ref{RSBhamil}),  and assume 
that the Gibbs state at $\lambda=0$ has no other symmetry. 
We say that a spontaneous RSB does not occur if 
 the following  two limiting procedures  commute for any  RSB  perturbation operator  ${\cal R}$ whose expectation exists in the infinite-volume limit for  
almost all  coupling constant $\lambda \in {\mathbb R}$, 
\begin{equation}
\lim_{N\to\infty } \lim_{\lambda \rightarrow  0}  {\mathbb E}  
\langle {\cal R}  \rangle_N,_\lambda = \lim_{\lambda \rightarrow  0} \lim_{N\to\infty } {\mathbb E}  
\langle {\cal R}  \rangle_N,_\lambda.
\label{commute}
\end{equation}
 }
Then, the following corollary is obtained from Theorem \ref{MT}. 

{\corollary \label{MT2}
Consider  a disordered quantum spin model defined by the Hamiltonian (\ref{RSBhamil})
 with a perturbation operator ${\cal O}_N$ defined by a RSB order operator ${\cal R =:O}_N$ satisfying Assumptions 1 and 2, and 
 assume that 
 the Hamiltonian (\ref{RSBhamil}) has no other symmetry at $\lambda=0$. 
 If spontaneous  replica symmetry-breaking (RSB) does not occur,  then
 RSB does not occur in Chatterjee's sense either.\\
 
 \noindent
Proof.}   
We prove the contrapositive. In the Hamiltonian (\ref{RSBhamil}),  
 Theorem\ref{MT} gives that
the infinite-volume limit of expectation value ${\mathbb E}\langle {\cal R} \rangle_N,_\lambda$ exists, and 
 the variance of the perturbation operator  ${\cal R}$ 
  vanishes 
\begin{equation}
\lim_{N \to \infty }{\mathbb E}\langle ( {\cal R}  -{\mathbb E} \langle {\cal R}  \rangle_N,_\lambda )^2 
\rangle_N,_\lambda  =0,
\end{equation} 
for almost all $\lambda \in {\mathbb R}$.
If  spontaneous RSB does not occur, then 
the two limiting procedures for ${\mathbb E} \langle {\cal R} \rangle_N,_\lambda$ and ${\mathbb E} \langle {\cal R}^2
 \rangle_N,_\lambda$ commute, as remarked in Note \ref{sRSB}. Therefore,
\begin{eqnarray}
&&\lim_{N \to \infty }[{\mathbb E}\langle{\cal R}^2  \rangle_N,_{0}  -({\mathbb E} \langle {\cal R}  \rangle_N,_0   )^2 ] \nonumber
=
 \lim_{N \to \infty }\lim_{\lambda \to 0}[{\mathbb E} \langle {\cal R}^2 \rangle_N,_\lambda  
 - ({\mathbb E} \langle {\cal R} \rangle_N,_\lambda  )^2 ]
 \\&&
=
\lim_{\lambda \to 0} \lim_{N \to \infty }[{\mathbb E} \langle {\cal R}^2 \rangle_N,_\lambda
  - ({\mathbb E} \langle{\cal R}  \rangle_N,_\lambda    )^2 ]
=0,
\end{eqnarray} 
for almost all $\lambda \in {\mathbb R}$. Since ${\cal R}$ is arbitrary, RSB does not occurs  in Chatterjee's sense.
$\Box$  

Note that Assumption 1 can be proven in  several short-range interacting spin models
for only $A:=\{1\}$ in the definition (\ref{calR}) of ${\cal R}$.
Corollary \ref{MT2} is valid under Assumption 1.

\end{document}